\documentclass[aps,prb,twocolumn,showpacs,showkeys,amsmath]{revtex4}
\usepackage{graphicx}
\begin{document}

\newcommand{\nt}{N$\acute{\textrm{e}}$el temperature}

\title{Field induced phases in UPt$_2$Si$_2$}
\author{D. Schulze Grachtrup$^{1}$, M. Bleckmann$^{1,2}$, B. Willenberg$^1$, S. S\"ullow$^1$, M. Bartkowiak$^3$, Y. Skourski$^3$, H. Rakoto$^4$, I. Sheikin$^5$, J. A. Mydosh$^6$}
\affiliation{
$^1$Institut f\"{u}r Physik der Kondensierten Materie, TU Braunschweig, Braunschweig, Germany\\
$^2$Wehrwissenschaftliches Institut f\"{u}r Werk- und Betriebsstoffe, Erding, Germany\\
$^3$Hochfeld-Magnetlabor Dresden, Helmholtz Zentrum Dresden Rossendorf, Dresden, Germany\\
$^4$Laboratoire National des Champs Magn$\acute{\textrm{e}}$tiques Intenses, Toulouse, France\\
$^5$Laboratoire National des Champs Magn$\acute{\textrm{e}}$tiques Intenses, Grenoble, France\\
$^6$Kamerlingh Onnes Laboratory, Leiden University, 2300RA Leiden, The Netherlands}

\date{\today}

\begin{abstract}
The tetragonal compound UPt$_2$Si$_2$ has been characterised as a moderately mass enhanced system with an antiferromagnetic ground state below $T_N$= 32 K. Here, we present an extensive study of the behavior in high magnetic fields. We have performed pulsed field magnetization and static field resistivity measurements on single crystalline samples UPt$_2$Si$_2$. Along the crystallographic $a$ axis, at low temperatures, we find a metamagnetic-like transition in fields of the order 40 T, possibly indicating a first order transition. Along the crystallographic $c$ axis, in magnetic fields of $B$ $\geq$ 24~T, we find distinct anomalies in both properties. From our analysis of the data we can distinguish new high field phases above the AFM ground state. We discuss the emergence of these new phases in the context of Fermi surface effects and the possible occurrence of a Lifshitz or electronic topological transition, this in contrast to previous modellings of UPt$_2$Si$_2$ based on crystal electric field effects.
\end{abstract}



\maketitle

\section{Introduction}
Throughout the last decades, ternary intermetallic uranium compounds of composition U$T_2$$M_2$, with $T$ a transition metal and $M$ either Si or Ge, stood in the focus of intensive research efforts. In this field, there are various topics that are addressed in experimental and theoretical studies. For instance, the heavy fermion superconductor URu$_2$Si$_2$ \cite{palstra} exemplifies the topic "hidden order" \cite{amitsuka1}, which presently is dicussed in terms of spin nematic phases \cite{okazaki,fujimoto}. The very same material is also a prime example for the observation of exotic field induced phases \cite{harrison,suslov,kim}, and whose microscopic nature have not been resolved so far. The series of materials U$T_2$Si$_2$ has served as testing ground for advanced band structure calculations \cite{sandratskii}, with experimental tests from many different authors (see for instance Refs. \cite{rebelsky,matsuda1,matsuda2,morkowski,svoboda,plackowski,endstra} and references therein). URh$_2$Ge$_2$ and UPt$_2$Si$_2$ have served as model materials to study disorder effects in correlated electron materials \cite{sullow1,sullow2,sullow3,sullow4}. Especially for the latter compound, UPt$_2$Si$_2$, the reinvestigations in the Refs. \cite{sullow4,ikeda,bleckmann} brought up the issue of the degree of $f$ electron localization in U compounds, a topic also discussed in context with other uranium intermetallics such as the heavy fermion superconductors UPd$_2$Al$_3$ or UPt$_3$ \cite{zwicknagl1,dressel,zwicknagl2,fujimori,rueff,mcmullan}.

For UPd$_2$Al$_3$ already a short time after discovery of the system the concept of different degrees of localization of the $f$ electron system was formulated \cite{caspary,feyerherm}. In contrast, UPt$_2$Si$_2$ was considered to be one of the rare examples of uranium intermetallics with strongly localized $f$ electrons, even allowing to describe the material properties based on a well defined crystal field scheme \cite{nieuwenhuys,steeman,amitsuka2}. This way, it was accounted for the temperature dependence of the susceptibility and the high field magnetization at lowest temperatures. Recently, in a brief report on a reinvestigation of the high field magnetization we have presented arguments against this interpretation \cite{schulze}. Here, we now present a full account of our high field studies, this way adding to our arguments in favour of an alternative view of the properties of UPt$_2$Si$_2$. In particular, we report on the discovery of new field induced phases in UPt$_2$Si$_2$, which we tentatively associate to Lifshitz type transitions. Altogether, we believe that with the experimental properties presented so far UPt$_2$Si$_2$ is a candidate material to be studied by advanced band structure calculations, to verify or disprove the phenomenological descriptions put forth by experimentalists.

UPt$_2$Si$_2$ was shown to crystallize in the tetragonal CaBe$_2$Ge$_2$ structure (space group P4/$nmm$) \cite{hiebljnm}. Below $T_N$ = 32~K the system orders antiferromagnetically with ferromagnetically coupled layers in the $ab$ plane, antiferromagnetic stacking of this layers along the $c$ axis and a moment of about 2 $\mu_B$ per uranium atom at 4.2 K pointing into the $c$ direction~\cite{sullow4,steeman,ptas,steemanjmmm}. The linear term of the specific heat was reported to $\gamma$ = 32 mJ mole$^{-1}$ K$^{-2}$ indicating a moderate mass enhancement of the electrons~\cite{steeman}.

Various physical properties of UPt$_2$Si$_2$ in low magnetic fields, such as susceptibility, magnetization and specific heat, were described in terms of a crystal electric field scheme within mean field approximation even though the calculated magnetic moment of $\approx$ 2.9 $\mu_B$ per uranium atom in this model is larger than the experimentally observed one \cite{nieuwenhuys,sullow4,steeman,ptas,steemanjmmm}. Further, in high field magnetization measurements carried out by Amitsuka \textit{et al.} \cite{amitsuka2}, besides of the discrepancies in the magnitude of the magnetic moment, additional multistep-like fine structures together with hysteresis were observed. Still, the authors concluded that the CEF model qualitatively is correct and that the observed discrepancies could result from not taking into account hybridisation effects between $5f$ and conduction electrons in the CEF modelling \cite{amitsuka2}. In contrast, based on new high field magnetisation data, we have concluded that these data do not support a CEF modelling of UPt$_2$Si$_2$ \cite{schulze}. 

Recently, UPt$_2$Si$_2$ was reexamined in the context of disorder affecting the properties of $f$ electron intermetallics \cite{sullow4}. Here, it was demonstrated that the electronic transport along the $c$ axis can be described by localization theory at temperatures $T > T_N$, while electron-magnon scattering contributes in the antiferromagnetic phase. At $T_N$, a Fermi surface reconstruction, indicated by a maximum of $\rho(T)$ within the ordered phase for the $c$ axis data, occurs. From a comparison of resistivity along the $c$ axis and neutron diffraction the transition temperature was identified in the resistivity as a minimum of $\partial \rho / \partial T$~\cite{sullow4}. Along the $a$ axis the resistivity is that of a common antiferromagnetic uranium intermetallic with moderate mass enhancement.

In the following, we present a full account of our high field study by discussing field and temperature dependent magnetization and resistivity experiments. From our data we extract the magnetic phase diagram for fields applied along the crystallographic $a$ and $c$ axes. We finish by discussing on a phenomenological level possible scenarios to account for the observed physical properties.

\section{Experimental setup}

All measurements were performed on single crystalline UPt$_2$Si$_2$ grown in Leiden using a modified Czochralski method as described in Ref.~\onlinecite{menovsky}. The samples are bar shaped with a cross section of 1x1 mm$^2$ and a length of 5 to 10 mm, with the long side cut parallel either to the $a$ or the $c$ axis. The crystals have been characterized previously in zero magnetic field with neutron diffraction, susceptibility, specific heat, Hall and resistivity measurements (see Refs.~\cite{sullow4,bleckmann}). For magnetization measurements as-cast crystals were used, while the resistivity measurements were done using crystals that have been annealed for one week at 900$^\circ$C. We have previously shown~\cite{bleckmann} that the annealing process does not significantly alter the physical properties of UPt$_2$Si$_2$, hence, in terms of the ground-state properties and the behavior in magnetic fields as-cast and annealed material will give the same results.

The high field magnetization measurements were carried out in Toulouse at the Laboratoire National des Champs Magn$\acute{\textrm{e}}$tiques Intenses (LNCMI) as well as the Dresden High Magnetic field laboratory (HLD) in pulsed magnetic fields up to 53 T. The data were taken with pulsed field magnetometers recording the induced voltages of a coil surrounding the samples (for details see Ref.~\onlinecite{hld-magnets}) and corrected for additional magnetization contributions from the sample environment inside the pick-up coils.

Resistivity data in magnetic fields up to 9 T directed along the crystallographic $a$ and $c$ axes (longitudinal geometry) and down to 1.8 K were taken in Braunschweig using a standard four-wire ac-technique. Moreover, resistivity measurements in high static magnetic fields up to 28 T for the same geometry were performed at the Grenoble site of the LNCMI, here using a four-wire lock-in technique.

\section{Magnetization}

\begin{figure}[!ht]
\begin{center}
\includegraphics[width=0.9\linewidth]{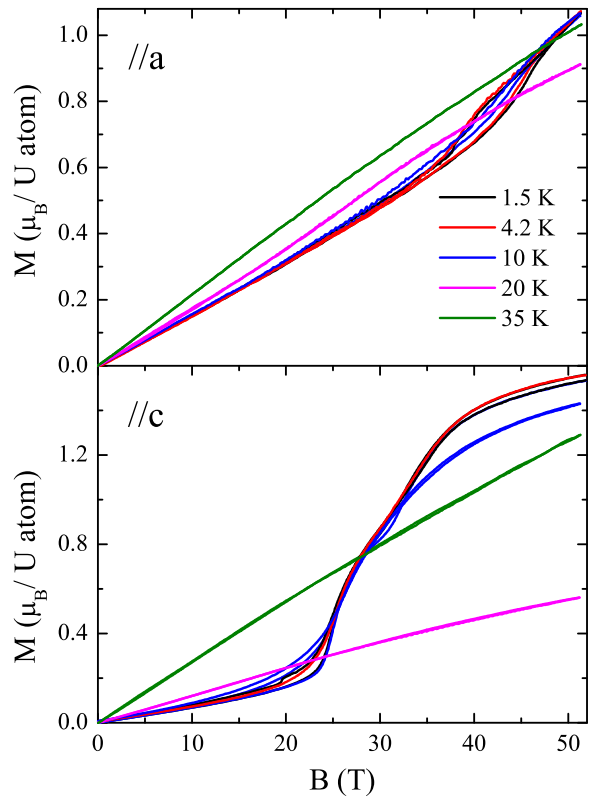}
\end{center}
\caption{(Color online) The magnetization $M(B)$ for $a$ and $c$ axes of single crystalline UPt$_2$Si$_2$ from measurements in pulsed magnetic fields. Legend applies to both graphs.}
\label{fig:MvsB}
\end{figure}

\begin{figure}[!ht]
\begin{center}
\includegraphics[width=0.9\linewidth]{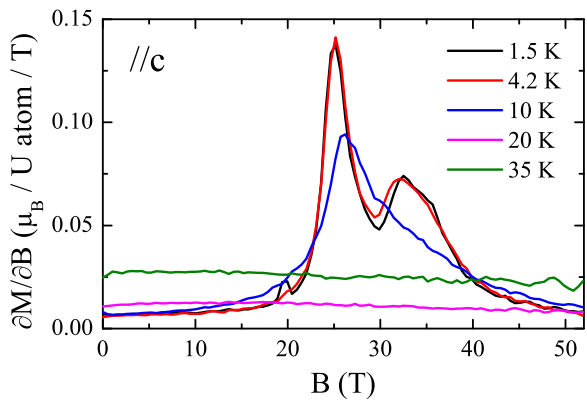}
\end{center}
\caption{(Color online) The field derivative of the magnetization $\partial M / \partial B$ of UPt$_2$Si$_2$ along the $c$ axis as function of the magnetic field $B$.}
\label{fig:dmdb}
\end{figure}

Our data from magnetization measurements at the LNCMI and the HLD cover the temperature range from 1.5 to 35 K, {\it i.e.}, from the AFM ordered phase in zero magnetic field into the paramagnetic regime above $T_N$. Both data sets contain the same significant field dependent features. Furthermore, at lowest temperatures the data agree well with low temperature data from Amitsuka {\it et al.}~\cite{amitsuka2}. In the Figs. \ref{fig:MvsB} and \ref{fig:dmdb} we show the data taken at the LNCMI, which cover a broader temperature range than the HLD data.

First, in Fig. \ref{fig:MvsB} the absolute magnetization $M$ for both crystallographic axes is depicted. Further, Fig. \ref{fig:dmdb} contains the field derivative $\partial M / \partial B$ for the $c$ axis to emphasize the most important features along this direction.

Along the $a$ axis at a high temperature of 35 K ({\it i.e.}, above the \nt), we see an almost linear dependence of $M(B)$ up to the highest fields, as one expects for an unsaturated paramagnet. On reducing temperature to 20 K, thus in the AFM phase, the magnetization is slightly reduced but qualitatively unchanged. In contrast, upon further reduction of temperature a metamagnetic-like transition appears, which in addition exhibits hysteresis in the field range from $\approx$ 35 to 48~T. Measurements at additional temperatures reveal that this distinct change of the magnetic behavior sets in between 16 and 20 K.

Along the $c$ axis at temperatures above $T_N$ again we find the expected almost linear $M(B)$ dependence up to the highest fields. However, in contrast to the $a$ axis data, for the $c$ axis there is now a strong reduction of the absolute magnetization at 52 T as the temperature is lowered from 35 K (1.28 $\mu_B /\text{U atom}$) to below $T_N$ (here 20 K: 0.56 $\mu_B /\text{U atom}$). Except for this difference in absolute values, both curves are qualitatively similar. This is also seen in Fig. \ref{fig:dmdb} with the derivatives $\partial M / \partial B$ at 20 and 35 K being almost constant but different in absolute numbers.

Upon further reduction of the temperature a very pronounced metamagnetic-like transition appears around 25 T at 10 K, indicating a qualitative difference between the magnetization curves at 10 and 20 K. This difference is also observed in the derivative $\partial M / \partial B$ (Fig. \ref{fig:dmdb}), where now a single peak appears. In addition, this metamagnetic transition is accompanied by weak hysteresis in the field range up to $\approx$ 25 T.

Finally, at even lower temperatures (here 4.2 and 1.5 K) a second metamagnetic-like transition appears, and which again is accompanied by weak hysteresis (for instance, at 1.5 K between $\approx$ 28 and 33 T). The occurrence of a new field induced phase is also reflected in $\partial M/\partial B$, where now two separate peaks are observed (Fig. \ref{fig:dmdb}).

\begin{figure}[!ht]
\begin{center}
\includegraphics[width=0.9\linewidth]{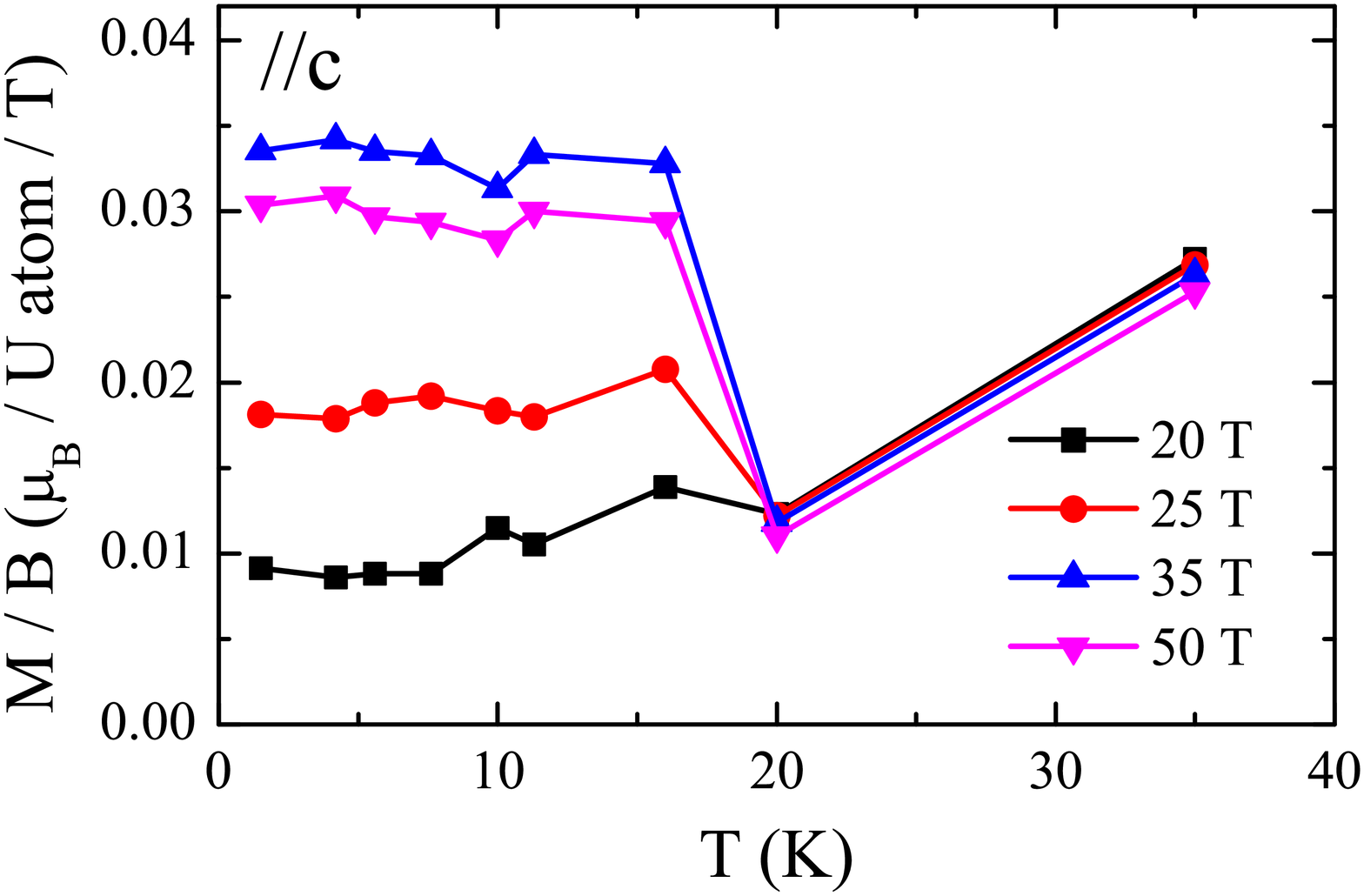}
\end{center}
\caption{(Color online) The susceptibility $M/B$ of UPt$_2$Si$_2$ along the $c$ axis as function of temperature for different magnetic fields. Data points are extracted from $M(B)$ measurements as depicted in Fig. \ref{fig:MvsB}.}
\label{fig:MvsT}
\end{figure}

A plot of the susceptibility $M/B$ (derived from the field dependence of the magnetization data) along the $c$ axis as depicted in Fig. \ref{fig:MvsT} highlights this temperature dependence in high magnetic fields. Up to about 20 T there is a slow monotonic increase with temperature from 1.5 to 35 K. In higher fields of  $B \sim $ 25 T and temperatures $\alt$ 16 K the susceptibility is almost constant, but with increased absolute values as compared to 20 T. Between 16 and 20 K at 25 T there is a sudden drop to the low field (20 T) values, indicating a qualitative change in behavior. A similar temperature dependence as in 25 T is observed for an external field of 35 T, but now again with an increase of the absolute values $M/B$ at temperatures $\alt$ 16 K. Further increase of the magnetic field (50 T) does not change absolute susceptibility values within experimental error.

Summarizing our findings, from the magnetization data along the $a$ axis we find a regime with a qualitative change of behavior for fields of the order 40 T at temperatures below $\sim$ 20 K. Furthermore, along the $c$ axis there is evidence for multiple field induced phases. The magnetization (Fig. \ref{fig:MvsB}) and field dependent susceptibility (Fig. \ref{fig:dmdb}) attest to phase transitions at around 25 T for temperatures below 20 K, and around 30 T below 10 K. The temperature dependent high field susceptibility (Fig. \ref{fig:MvsT}) points to phase transitions upon lowering temperature to $<$ 20 K in fields $\agt$ 25 T.

\section{Resistivity}

\begin{figure}[!ht]
\begin{center}
\includegraphics[width=0.9\linewidth]{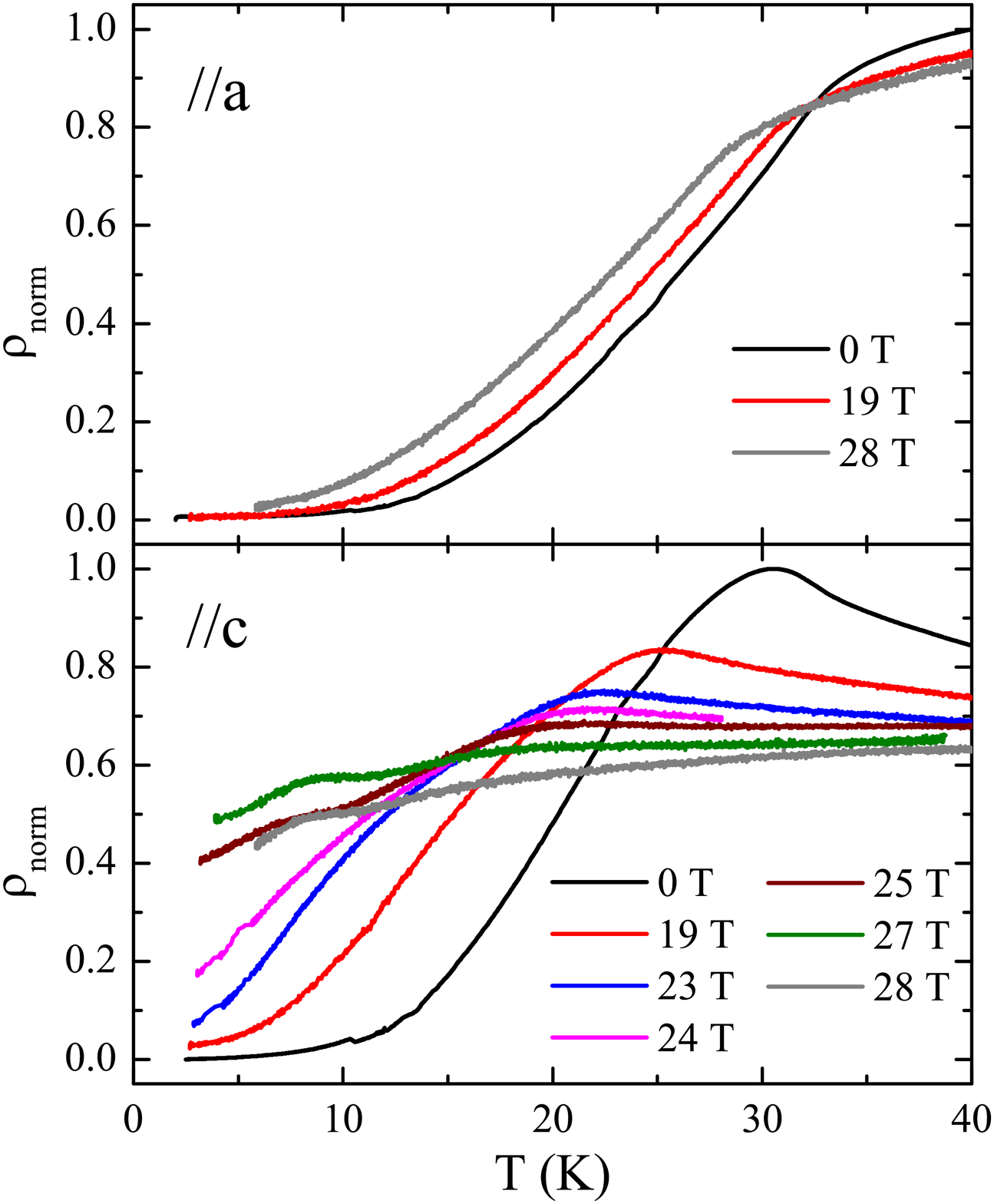}
\end{center}
\caption{(Color online) The normalized resistivity $\rho_{norm}(T)$ of single crystalline UPt$_2$Si$_2$ for selected fields from measurements in static magnetic fields along $a$ and $c$ axes as determined using Eq. \ref{eq:norm}.}
\label{fig:RvsT}
\end{figure}

Our resistivity data consists of multiple field and temperature sweeps covering the temperature range from 2 up to 40 K in magnetic fields up to 28 T. Temperature dependent measurements in zero magnetic field nicely reproduce all resistive features previously reported~\cite{steeman,sullow4,johannsen}. Due to the intrinsic disorder in UPt$_2$Si$_2$ the absolute values of the resistivity are strongly dependent on the sample used and the actual contact positions of the connecting wires~\cite{sullow4}. Therefore, even different measurements of the same crystal in the same setup yield different resistivities when the contacts are renewed between two measurements. Due to this variation of absolute resistivities we have normalized our experimental data of the temperature dependent measurements according to
\begin{equation}
\rho_{norm}=\frac{\rho(T)-\rho_{min}}{\rho_{max}-\rho_{min}},
\label{eq:norm}
\end{equation}
with $\rho_{min}$ and $\rho_{max}$ being the minimum and maximum resistivities in the temperature range 2 to 40 K in zero magnetic field. This normalization was carried out separately for each data set using the corresponding zero field values with the same contact positions. This way, field dependent effects are emphasized in our graphic representations, while the residual resistivity as the most important resistive component independent of field is corrected for in the data. In Fig. \ref{fig:RvsT} we show the resulting data for selected fields after the normalization procedure has been performed. 

For measurements with the field applied along the $a$ axis, the \nt ~$T_N$ shifts down to lower temperatures with increasing magnetic field, as is typical for the field response of an antiferromagnet. Based on a comparison of in-field with zero field data we find that the \nt ~can be identified as kink in $\rho_{norm}(T)$, {\it i.e.}, $\left| \partial^2 \rho_{norm}(T) / \partial T^2 \right| = max$. We note that the highest experimentally accessible field of 28 T was not sufficient to reach the field region where a new hysteretic metamagnetic transition appeared in the magnetization. 

In contrast to the $a$ axis, along the $c$ axis the antiferromagnetic phase transition at the \nt ~in zero magnetic field can be identified as the minimum in $\partial \rho / \partial T$ (see Ref. \cite{sullow4}). This minimum is a result of an anomalous and instantaneous increase of the resistivity upon entering the AFM ordered phase due to a Fermi surface reconstruction. Upon increasing the magnetic field this resisitive anomaly, and correspondingly the minimum in $\partial \rho / \partial T$, become less pronounced. Nevertheless traces of the anomaly indicative of the AFM phase transition are visible up to 27 T, with the \nt ~being shifted from 32 K in zero field to 23 K at 27 T.

Furthermore, starting at zero magnetic field and at temperatures below $\approx$ 13 K, the resistivity first increases with increasing field without qualitative changes up to 24 T. But in higher magnetic fields (here: at and above 25 T) a new and distinct anomaly appears, and which is clearly visible as an additional contribution to the resistivity. The magnitude of this anomaly reaches its maximum at 27 T and decreases upon further increase of the magnetic field reaching the 25 T value at 28 T again. This anomaly is indicative of another phase transition at a critical temperature $T^*$, which we determine from the condition $\left| \partial^2 \rho_{norm}(T) / \partial T^2 \right| = max$ and varies between 6.9 and 8.2 K in the field range 25 to 28 T.

\begin{figure}[!ht]
\begin{center}
\includegraphics[width=0.9\linewidth]{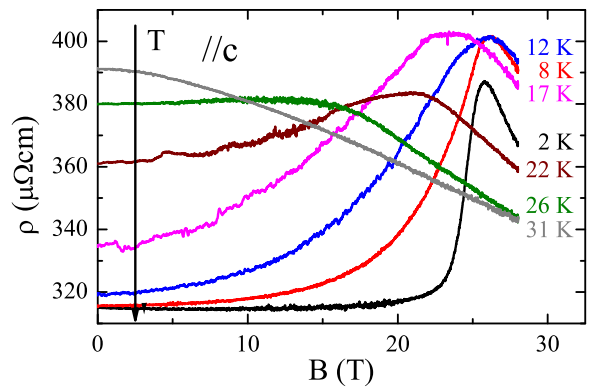}
\end{center}
\caption{(Color online) The absolute magnetoresistivity of single crystalline UPt$_2$Si$_2$ along the $c$ axis for selected temperatures from measurements in static magnetic fields.}
\label{fig:RvsB}
\end{figure}

This additional anomaly also leaves a trace in the field dependent resistivity along the $c$ axis, which is depicted as absolute magnetoresistivity in Fig. \ref{fig:RvsB}. At 2 K we first see an almost field independent resistivity up to $\approx$ 20 T. In higher magnetic fields the resistivity increases, with a peak-shaped field dependence reaching its maximum at 26 T. With increasing temperature this peak is strongly broadened while the overall curve transforms into a monotonously decreasing one at 31 K, close to the ordering temperature.

The mechanisms for the temperature dependence of the resistivity of magnetic materials such as UPt$_2$Si$_2$ at low $T$ are electron-electron $\rho_{el-el}$, electron-magnon $\rho_{el-mag}$ and electron-phonon scattering $\rho_{el-ph}$,
\begin{equation}
\rho(T)=\rho_0+\rho_{el-el}+\rho_{el-mag}+\rho_{el-ph},
\label{eq:rhomech}
\end{equation}
with $\rho_0$ as the residual resistivity. Especially, for the $c$ axis data on UPt$_2$Si$_2$ we have demonstrated that electon-phonon scattering can be neglected to first approximation, thus leaving electron-electron and electron-magnon scattering in the resistivity to be accounted for as described in Eq. \ref{eq:rho} (Ref. \cite{anderson}): 
\begin{equation}
\rho(T)=\rho_0+A T^2+\frac{DT}{\Delta}\left( 1+2\frac{T}{\Delta} \right) e^{-\frac{\Delta}{T}}.
\label{eq:rho}
\end{equation}
Here, $AT^2$ reflects electron-electron-scattering, while the exponential term covers scattering due to magnons which are excited across a spin wave excitation gap $\Delta$. 

We have used this function to fit our experimental data along the $c$ axis in the range up to $\frac{2}{3} \times T_N$ for fields up to 9 T, and up to $\frac{1}{3} \times T_N$ for higher fields, with the upper temperature limit variable to account for the decrease of $T_N$ with magnetic field. Although Eq. (\ref{eq:rho}) a priori is only valid in the low temperature limit ($T \ll T_N$), the fitted parameters do not change significantly upon increasing the temperature range of the fit while the accuracy clearly is improved. Thus, for UPt$_2$Si$_2$ Eq. (\ref{eq:rho}) can also be applied at higher temperatures.

\begin{figure}[!ht]
\begin{center}
\includegraphics[width=0.9\linewidth]{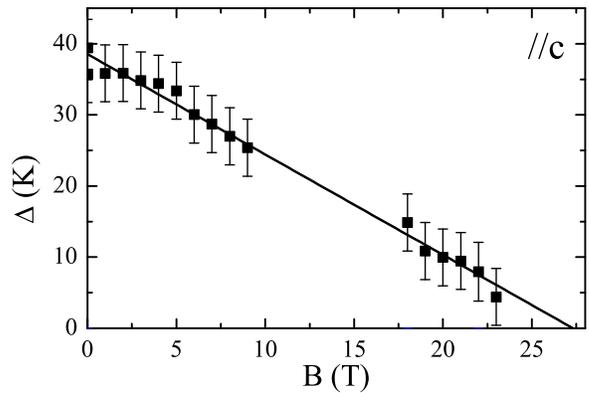}
\end{center}
\caption{The temperature dependence of the spin wave excitation gap $\Delta$ of UPt$_2$Si$_2$ from fits of Eq. \ref{eq:rho} to the experimental resistivity data along the crystallographic $c$ axis. The line is a guide to the eye and indicates the disappearance of the gap around 27 T.}
\label{fig:delta}
\end{figure}

From our fits we observe that the free parameter $A$ is quite small, which is consistent with the moderate mass enhancement in the specific heat ($A \propto \gamma^2$). In effect, electron-electron scattering accounts only for less then 10\% of the temperature dependence of $\rho(T)$. Therefore, we also approximate this parameter by fixing it at zero, a condition that does not alter the other fit parameters significantly. Also, this assumption is consistent with our data analysis previously carried out in zero magnetic field, where the resistive contributions from electron-electron scattering turned out to be rather small~\cite{sullow4}. Correspondingly, from our fits we obtain the field dependence of the spin wave excitation gap $\Delta$, which is depicted in Fig. \ref{fig:delta}. In these fits, the prefactor $D$ also exhibits a field dependence, so fit parameter interdependency might be an issue here, and which gives rise to some uncertainty in the determination of $\Delta$ as indicated in the plot by error bars.

Keeping in mind this uncertainty, our analysis essentially yields a suppression of the spin wave excitation gap $\Delta (B)$ with applied magnetic field. From Fig. \ref{fig:delta} we conclude that the spin wave excitation gap closes at around 27 T. In other words, the field dependent resistivity data $\rho (T)$ along the $c$ axis of UPt$_2$Si$_2$ suggests that the zero field spin excitation gap $\Delta$ dissappears in that field range, where from the resistivity and magnetization we have established the appearance of new field induced phases.  

The analysis of the $a$ axis data is more intricate. In Ref.~\cite{sullow4} we have demonstrated that for both $a$ and $c$ axes the resistivity in the AFM state can be described using Eq. \ref{eq:rho}. However, since there is no maximum in $\rho (T)$ just below $T_N$ for the resistivity along the $a$ axis, a priori there is no fundamental argument to prove that this assumption is correct. Alternatively, it could be assumed that the temperature dependence of $\rho (T)$ below $T_N$ is the result of electron-magnon scattering without excitation gap, and which is parameterized by (neglecting again phonon scattering)
\begin{equation}
\rho(T)=\rho_0+A T^2+B T^x.
\label{eq:rhomag}
\end{equation}
Here, $x$ is an exponent depending on the type of magnetic fluctuations (for instance, according to Ref. \cite{anderson}, $x = 2$ for ferromagnetic and $x = 5$ for antiferromagnetic fluctuations). In this situation, for the $a$ axis data we have carried out two types of analysis, first by fitting the data using Eq. \ref{eq:rho}, secondly by parameterizing our data utilizing Eq. \ref{eq:rhomag}. Both types of fits reproduce the experimental data equally well, therefore we cannot draw definite conclusions about the possibility of a field dependence of the various parameters in the Eqs. \ref{eq:rho} or \ref{eq:rhomag}.

\section{Discussion}
\label{sec:diss}

\begin{figure}[ht]
\begin{center}
\includegraphics[width=0.9\linewidth]{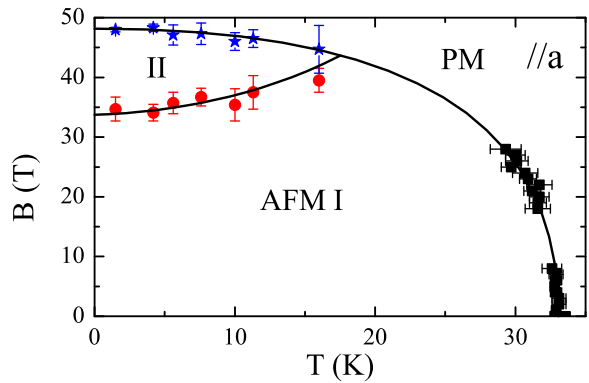}
\end{center}
\caption{(Color online) Magnetic phase diagram of UPt$_2$Si$_2$ for magnetic fields applied along the hard magnetic $a$ axis. Black squares indicate $T_N$ as obtained from the maximum curvature of $\rho(T)$, while red circles and blue stars denote the lower ($B_{a1}$) and upper ($B_{a2}$) boundary of the hysteretic area. Solid lines schematically indicate phase boundaries, for details see text.}
\label{fig:phasesa}
\end{figure}

Based on the magnetization and resistivity data presented we can now proceed and construct the magnetic phase diagrams for fields aligned along both $a$ and $c$ axes. The data analysis for the $a$ axis is straightforward and based upon ({\it i.}) the data $\rho (T)$ up to 28 T and ({\it ii.}) the data $M (B)$ up to 52 T. From the resistivity we find a continuous suppression of $T_N$ with field $B$. Further, from the magnetization we can extract the fields $B_{a1}$ and $B_{a2}$ of start and end of the hysteretic regime by determining the fields at which the $M(B)$ data for the field-sweep up and field-sweep down deviate from and join each other, respectively. Using the values $T_N (B)$, $B_{a1}$ and $B_{a2}$, we arrive at the phase diagram for the $a$ axis as depicted in Fig. \ref{fig:phasesa}.

Altogether, along the $a$ axis we see a smoothly decreasing \nt ~ defining the AFM phase I with increasing magnetic field up to 28 T, implying that we have the typical phase diagram for an antiferromagnet down to $\approx$ 18 K. Then, at lower temperatures between 16 K and 20 K, we observe a distinct change in the magnetization with the appearance of a metamagnetic-like transition and hysteresis between field-sweep up und down. Here, the characteristic fields $B_{a1}$ and $B_{a2}$ define the hysteretic region II. The observation of magnetic hysteresis possibly might indicate that the second order AFM phase transition at high temperatures becomes first order at low temperatures/high fields, as it has been observed and calculated previously \cite{schmidt,stryjewski,katsumata,kincaid, selke, santos, zukovic}.

At first glance, the phase diagram along the $a$ axis bears close resemblance to those of other antiferromagnets \cite{stryjewski} such as FeCl$_2$, FeBr$_2$ or CoCl$_2\cdot$2H$_2$O. In those materials the phase transitions occur from a ground state in which the magnetic field is applied parallel to the easy magnetic axis. For UPt$_2$Si$_2$, however, this condition is not fulfilled, since neutron scattering experiments prove the $c$ axis to be the easy axis \cite{steeman,sullow5}. To our knowledge in the literature there are no materials or models which exhibit similar characteristics for a geometry of the magnetic field applied along the hard magnetic, in our case the $a$ axis as in UPt$_2$Si$_2$.

There have been efforts to calculate the magnetic phase diagrams of an Ising metamagnet in combined longitudinal and transverse fields \cite{wei}. These calculations reproduce some features of the phase diagram along the $a$ axis, especially a crossover from a first-to-second order phase transition along the magnetically hard axis. However, such a crossover only occurs in a combination of sufficiently large transverse and longitudinal fields. This is not the case for our measurements on UPt$_2$Si$_2$. Here, longitudinal fields along the $a$ axis would only arise from imperfect alignment of the crystal and will be much smaller ($<5^\textrm{o}$) than discussed in Ref. \cite{wei}, thus this model cannot account for the observed properties of UPt$_2$Si$_2$.

As a last line of thought, we note that for the related uranium intermetallic UPd$_2$Al$_3$ a hysteretic metamagnetic high field transition has been observed \cite{sullow6,oda,devisser}. As yet, a conclusive explanation for such behavior has not been put forth, although it might be speculated that the hysteretic behavior is a result of significant magnetoelastic coupling, thus reflecting a structural response of the system onto a change of the magnetic state. Analogously, it might be argued that for UPt$_2$Si$_2$ the observation of hysteresis in the magnetization $\| a$ axis does reflect a first order phase transition associated to magnetoelastic effects. Unfortunately, on behalf of theoretical modelling such issues have hardly been investigated for uranium intermetallics in recent years. 

\begin{figure}[ht]
\begin{center}
\includegraphics[width=0.9\linewidth]{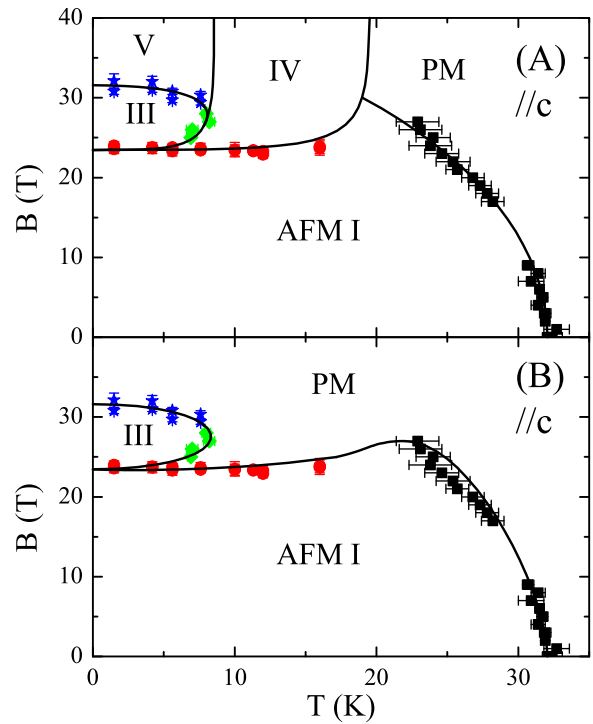}
\end{center}
\caption{(Color online) Suggested magnetic phase diagrams of UPt$_2$Si$_2$ for magnetic fields applied along the soft magnetic $c$ axis according to the scenarios (A) and (B) discussed in the text. Black squares and green diamonds define the transition temperatures $T_N$ and $T^*$ as obtained from resistivity measurements in magnetic fields. The red circles and blue stars indicate the critical fields $B_{c1}$ and $B_{c2}$ of metamagnetic-like transitions as obtained from the maxima in $\partial M / \partial B$. Solid lines schematically indicate phase boundaries, for details see text.}
\label{fig:phasesc}
\end{figure}

Constructing the magnetic phase diagram for the field $B$ applied along the $c$ axis is a much more complicated task. In a first step, we determine the phase boundary at $T_N$ between paramagnetic and low field antiferromagnetic phase I using the resistivity measurements in magnetic field from the condition $\partial \rho / \partial T = min$ (black squares in Fig. \ref{fig:phasesc}). As for the $a$ axis, we observe the common behavior of an antiferromagnet with a smooth decrease of $T_N$ down to $\sim$ 23 K with field $B$ up to 27~T. Secondly, from the magnetization $M (B)$ at different temperatures we can extract a first critical field $B_{c1}$ of a metamagnetic-like transition by determining the (first) maximum in the field derivative $\partial M / \partial B$ (see Fig. \ref{fig:dmdb}), this way defining a phase boundary as indicated by the red circles in Fig. \ref{fig:phasesc}. A third set of phase boundary data we obtain from the high field ($\geq 25$ T) resistivity, which as function of temperature shows two transitions. As described above (see Fig. \ref{fig:RvsT}), this second transition temperature $T^*$ is obtained from the condition $\partial^2 \rho_{norm}(T) / \partial T^2  = min$ (green diamonds in Fig.~\ref{fig:phasesc}). Finally, for the magnetization $M (B)$ at $T \leq 8$ K a second metamagnetic-like transition occurs, with a critical field $B_{c2}$ determined from the position of the second maximum in $\partial M / \partial B$ in Fig. \ref{fig:dmdb} (blue stars in Fig.~\ref{fig:phasesc}).

In the magnetic field range $B >$ 28 T only magnetization data is available, leading to ambiguities in defining phase boundary lines. Therefore, we will discuss two different scenarios for the magnetic phase diagram along the $c$ axis: First we will present scenario (A) with multiple distinct high field phases as depicted in Fig. \ref{fig:phasesc}(A). Subsequently a second scenario (B) as depicted in Fig. \ref{fig:phasesc}(B) with only one enclosed high field phase III and a polarized paramagnetic phase at low temperatures and high fields will be presented.

Both phase diagrams (Figs. \ref{fig:phasesc}(A) and \ref{fig:phasesc}(B)) have common phase boundarie lines. With the data for the phase diagram $B \| c$ axis put together as described above, we can now draw these lines. First of all, there is the boundary between paramagnetism and AFM phase I at comparatively low fields. Secondly, the critical field $B_{c1}$ separates the AFM phase I from a distinct high field state. And thirdly, the data points $B_{c1}$, $B_{c2}$ and $T^*$ indicate that there is a distinct (antiferromagnetic?) phase III in the field range 24 to 32 T at temperatures $T \leq 8$ K.

We proceed by discussing scenario (A) with multiple distinct high field phases and construct the corresponding phase digram (Fig. \ref{fig:phasesc}(A)). The temperature dependent susceptibility in Fig. \ref{fig:MvsT} indicates that by changing temperature from 1.5 to 35K in 20 T only the boundary from phase I to paramagnetism is crossed. In contrast, in a field of 25 T there is the pronounced dip between 16 and 20 K, suggesting that the phase boundary defined by $B_{c1}$ exhibits an upward curvature as indicated by the corresponding line in Fig. \ref{fig:phasesc}(A). In other words, in 25 T at 20 K the system resides in the AFM phase I, while in this field at lower temperatures - say at 16 K - there is a new and distinct phase IV. For higher fields (here 35 and 50 T), the temperature dependent susceptibility in Fig. \ref{fig:MvsT} suggests that there is a phase boundary between 16 and 20 K, suggesting an evolution of the boundary line as depicted in Fig. \ref{fig:phasesc}(A). 

Finally, at temperatures $\leq 8$ K, there is a sequence of field induced transitions $B_{c1}$/$B_{c2}$ in the magnetization. It implies that the thermodynamic state at low temperature/high fields (say, at 35 T and 5 K) is different both from that in phase III as in phase IV. The latter conclusion can also be drawn by directly comparing the magnetization for $B \| c$ axis at 4.2 and 10 K in Fig. \ref{fig:MvsB}. Clearly, there is a qualitative difference in the field dependence of $M$ in high fields. This in turn requires yet another phase boundary to exist just below 10 K in high magnetic fields, and which separates phase IV from phase V. Schematically, we have included this phase boundary line in Fig. \ref{fig:phasesc}(A), assuming that the line must be almost vertical, since in our field dependent magnetization measurements we have no indication to cross this line. Altogether, we thus arrive at a very rich high field phase diagram for UPt$_2$Si$_2$ along the $c$ axis in this scenario (A).

The main concern regarding scenario (A) is that there is only circumstantial evidence for the existence of the boundary lines defining the phases IV and V, but no direct experimental evidence. Therefore we will also discuss an alternative scenario (B) as depicted in Fig. \ref{fig:phasesc}(B) without such almost vertical phase boundaries.

In scenario (B) the phase boundaries (i) separating the paramagnetic and the AFM I phase ($T_N$), (ii) separating the AFM I phase from the high field regime ($B_{c1}$) and (iii) enclosing phase III ($B_{c1}$, $B_{c2}$ and $T^*$) remain the same as in scenario (A). Now, from this starting point, the only possibility to connect both parts of the phase boundary of the AFM I phase (red points and black squares in Fig. \ref{fig:phasesc}(B)) without proposing additional phases requires a maximum to exist for this boundary line at $\sim$ 20 K and $\sim$ 24 T. Now in high magnetic fields $B \geq$~32~T there are no phase boundaries in this scenario implying that the paramagnetic phase at high temperatures and low magnetic fields becomes field polarized with increasing applied field and reduced temperature. Thus a crossover from a paramagnetic to a field polarized state with ferromagnetic character occurs.

Now, also for scenario (B) there are a few question marks about its validity. First, the second order phase transition from the paramagnetic to the antiferromagnetic phase in zero magnetic field (black squares in Fig. \ref{fig:phasesc}(B)) is directly connected to the hysteretic metamagnetic-like phase transition at $\sim$ 24 T (red circles in Fig. \ref{fig:phasesc}(B)). Such a direct connection of both phase boundaries requires either a first-to-second order crossover at $\sim$ 20 K or an explanation of the hysteretic magnetization together with a second order phase transition. Second, for a crossover regime to a field polarized paramagnetic state in high magnetic fields one would expect a smooth evolution of physical properties with decreasing temperature/increasing polarization. This, however, is in conflict with our measurements of $M/B$ (Fig. \ref{fig:MvsT}), where we find a field independent regime at $T \geq$ 20 K, a strong change between 16 and 20 K and a temperature independent value of $M/B$ for $T \leq$ 16 K. To resolve these issues and to verify if scenario (A) or (B) are appllicable to UPt$_2$Si$_2$, in the future additional high field studies will have to be carried out.

Given this ambiguity regarding phase boundary lines in UPt$_2$Si$_2$, one might even speculate that the field derivative of the magnetization $\partial M / \partial B$ along the $c$ axis (Fig.~\ref{fig:dmdb}) contains yet another additional high field transition. This line of thought would be motivated by the rather broad maximum of $\partial M / \partial B$ at 1.5 / 4.2 K in the range 30 to 40 T. It could be argued that the shoulder at 36 T indicates a distinct phase transition, which in some way might connect to one of the other transition lines. However a quantitative analysis of $\partial M / \partial B$ in this field range turns out to be an awkward procedure, hence we did not follow up on this idea. We note that the uncertainty regarding the evolution of phase boundary lines might also be a result of structural disorder, which is inherent to UPt$_2$Si$_2$. As an example, in a related heavy fermion system, UNi$_2$Al$_3$, it has been found that the observation of subtleties of the magnetic phase diagram depended on the sample quality \cite{sullow7,sato}. Such issues will be more prominent for an inherently disordered system such as UPt$_2$Si$_2$.

Other uranium and cerium based intermetallics, for instance URu$_2$Si$_2$, CeRu$_2$Si$_2$, UPt$_3$ and UPd$_2$Al$_3$, although they are discussed in a different physical context, show similarities with respect to novel field induced phases and metamagnetism \cite{harrison,kim,jaime,haen,paulsen,hasselbach,adenwalla,sullow6,oda,devisser}. Usually these systems are discussed in the context of heavy fermions with a strong itinerant character, while UPt$_2$Si$_2$ appears to be much more localized, therefore a direct comparison does not seem to be appropriate. Still, some of the physical properties and features of the phase diagrams resemble those of UPt$_2$Si$_2$, possibly indicating a link between the properties of those rather itinerant systems and UPt$_2$Si$_2$.

Conversely, a comparison with more localized materials such as the Ising-type metamagnet FeBr$_2$ might yield further insight. At first glance, there appear to be similarities between our data in Fig. \ref{fig:phasesc} and that of FeBr$_2$ \cite{azevedo}. Although this system belongs to a class of materials with different underlying physical mechanisms, the phase diagrams resemble each other. In FeBr$_2$ a paramagnetic (PM), an antiferromagnetic (AF) and a mixed phase (AF+PM) exist. If there were analogies between FeBr$_2$ and UPt$_2$Si$_2$, these phases would correspond to our phases PM, AFM I and III, with phase III being a mixed phase. For such a mixed phase the magnetization is expected to vary slowly with applied field due to the growth of the volume amount of the phase with higher magnetization at the cost of the second one. In addition, hysteresis should be observed in the whole field range of the mixed phase. Neither is observed in our experiments (Fig. \ref{fig:MvsB}), where two step-like metamagnetic transitions at the phase boundaries AFM I $\leftrightarrow$ III and III $\leftrightarrow$ V (scenario (A)) or III $\leftrightarrow$ PM (scenario (B)), together with hysteresis over a small field range occur. Therefore, we rule out an Ising-type scenario similar to that in FeBr$_2$ to account for our findings on UPt$_2$Si$_2$.

As already discussed for the $a$ axis above, the magnetic phase diagram of an Ising metamagnet in longitudinal and transverse fields has also been calculated for the easy magnetic (here $c$) axis \cite{wei}. Again, for appropriate parameters, a crossover from a first to a second order phase transition is predicted, together with a phase boundary resembling that of our AFM phase I along the $c$ axis. However, the calculations give no indication of additional high field phases as we have detected them above 24 T. Thus this model cannot account for all the observed properties of UPt$_2$Si$_2$.

Rather than using spin reorientation models to account for our observations in UPt$_2$Si$_2$, we attribute the observed properties to Fermi surface effects. This interpretation is supported by two facts: First, the distinct increase in $\rho$(T) just below $T_N$ along the $c$ axis indicates Fermi surface effects in UPt$_2$Si$_2$ to take place. As well, the second transition in fields $\sim$ 25 T $\|c$ axis observed in the resistivity, together with the signifcantly larger residual resistivity in this field range is indicative of Fermi surface effects. Secondly, as we have pointed out before, the high field magnetization cannot be accounted for by a localized crystal electric field scheme, but which instead must be treated in a more itinerant picture \cite{schulze}. Thus, Fermi surface effects play a key role in understanding the high field properties of UPt$_2$Si$_2$.

Phase transitions due to Fermi surface effects are well-known in the literature as Lifshitz or electronic topological transitions (ETT) \cite{lifshitz, blanter, varlamov, yamaji, andrianov}. Such transitions occur when electronic bands are shifted through the Fermi energy and the Fermi surface topology is changed upon variation of external parameters such as magnetic field or pressure. These changes could correspond for instance to a disruption of a neck of the Fermi surface or the creation of new holes inside a portion of the Fermi surface. At zero temperature, this leads to a singularity in the density of states and a singularity in the thermodynamic potential $\Omega \propto(E-E_F)^{5/2}$. As a result, thermodynamic and transport properties show singularities as well, and which are semeared out at finite temperatures \cite{lifshitz,blanter,varlamov}.

Clear experimental evidence for a Lifshitz transition is difficult to find due to the thermal broadening and other ambiguities in data interpretation. Therefore, typically the existence of a Lifshitz transition is established by a combined analysis of experimental data and band structure calculations. Although most transitions of this kind found so far occured in nonmagnetic materials, Lifshitz transitions have also been proposed to occur in magnetic materials recently \cite{zwicknagl3,stockert,andrianov,kozlova,daou,rourke,yelland}. Correspondingly, it seems possible that in UPt$_2$Si$_2$ such a transition might occur as well. In this case, we would expect that in advanced band structure calculations on UPt$_2$Si$_2$ distinct topological Fermi surface features close to the Fermi energy should exist, and which can be shifted through the Fermi energy by magnetic fields of the order 20 to 50 T. As a result such features might cause field induced phase transitions and produce corresponding phase boundaries as we have observed experimentally.

An alternative approach to account for our observations might be found in the concept of different degrees of $f$-electron localization proposed for some uranium heavy fermions \cite{zwicknagl1,zwicknagl2,zwicknagl4,dressel,caspary,feyerherm}. Conceptually, for UPt$_2$Si$_2$ we might use this approach to produce a mixture of localized and itinerant views of the electronic structure in correlated electron materials. One might imagine that in band structure calculations some of the $f$-electron orbitals could be described as being localized, others being itinerant. This way, it might be possible to reproduce the ambivalent localized/delocalized character of UPt$_2$Si$_2$.

Further experimental access to the new found field induced phases in UPt$_2$Si$_2$ as well as to gain further insight into the microscopy of these phases is difficult due to the high magnetic fields of up to 50 T involved. Such magnetic fields, with the fast changes in magnetic flux and the small time scales involved in pulsed field measurements, rule out many experimental techniques. Furthermore, quantum oscillation measurements via for instance the de Haas-van Alphen-effect, although usually being a technique applicable in high magnetic fields, can most likely not be carried out in UPt$_2$Si$_2$ due to the intrinsic disorder and the corresponding small electronic relaxation times which result in a small Dingle factor. Still, there might be possibilities to gain additional insight into the nature of the field induced phases. First, along the $c$ axis (Fig. \ref{fig:phasesc}) the phases III (and IV in scenario (A)) can be accessed by DC magnetic fields in certain laboratories. Here techniques like Hall effect, NMR or specific heat could possibly be used to gain information on the density of states close to the Fermi energy, on the local magnetic structure and on the first or second order nature of the of the observed transitions. Secondly, measurements of the ground state Fermi surface (in the AFM I phase), for instance with ARPES, might allow to observe structures close to the Fermi energy which might produce phase transitions as observed.

As conclusion we have performed high magnetic field magnetization and resistivity measurements in pulsed and static magnetic fields up to 52 T. From these measurements we derived the magnetic phase diagrams along the crystallographic $a$ and $c$ directions. A comparison to other materials and models discussed in the context of metamagnetism was carried out. A close inspection of apparent similarities ruled out that these materials and models could be used to explain the properties of UPt$_2$Si$_2$. Instead, we proposed Fermi surface effects to produce the field induced phases. Band structure calculations taking into account Lifshitz transitions or varying degrees of $f$-electron localization might help to resolve these issues.

\begin{acknowledgments}
Part of this work has been supported by the "Transnational Access Program - Contract nr. 228043 - Euromagnet II - Integrated Activities" of the European Commission. The samples have been produced within FOM/ALMOS. We acknowledge fruitful discussions with G. Zwicknagl.
\end{acknowledgments}


\end{document}